\documentclass [11pt,letterpaper]{article}
\pdfoutput=1

\usepackage{jheppub}
\usepackage[mathscr]{eucal}
\usepackage{bm}
\usepackage{epsfig}
\usepackage[latin1]{inputenc}
\usepackage{float}
\usepackage{graphicx}
\usepackage{cancel}
\usepackage{mathrsfs}
\usepackage{amssymb}
\usepackage{amsfonts}
\usepackage{amsmath}
\usepackage{slashed}
\usepackage{hyperref}

\usepackage{caption}

\title{Chiral drag force}

\author[a]{Krishna~Rajagopal,}

\author[a,b]{Andrey V. Sadofyev}
\affiliation[a]{Center for Theoretical Physics, Massachusetts Institute of Technology, Cambridge, MA 02139}
\affiliation[b]{ITEP, B. Cheremushkinskaya 25, Moscow, 117218 Russia}

\emailAdd{krishna@mit.edu}
\emailAdd{sadofyev@mit.edu}

\preprint{MIT-CTP-4676}

\abstract{We provide a holographic 
evaluation of novel contributions to the drag force acting 
on a heavy quark moving through strongly interacting plasma. 
The new contributions are chiral in the sense that they
act in opposite directions in plasmas containing an
excess of left- or right-handed quarks.
The new contributions are proportional
to the coefficient of the axial anomaly, and in this
sense also are chiral.  These new contributions to the drag
force act either parallel to or antiparallel to an external
magnetic field or to the vorticity of the fluid plasma.
In all these respects, these contributions to the
drag force felt by a heavy quark are analogous to
the chiral magnetic effect (CME) on light quarks.  However,
the new contribution to the drag force is independent of the
electric charge of the heavy quark and is the same for
heavy quarks and antiquarks, meaning that
these novel effects do not in fact contribute to the CME current.
We show that although the chiral drag force can be non-vanishing
for heavy quarks that are at rest in the local fluid rest frame,
it does vanish for heavy quarks that are at rest in a 
suitably chosen frame.
In this frame, the heavy quark at rest sees counterpropagating
momentum and charge currents, both proportional
to the axial anomaly coefficient, but feels no drag force.
This 
provides strong concrete evidence
for the absence of dissipation in 
chiral transport, something
that has been predicted previously via consideration of symmetries. 
Along the way to our principal results, we provide a general calculation
of the corrections to the drag force due to the presence of 
gradients in the flowing fluid in the presence of a nonzero chemical potential.
We close with a consequence of our result that is at least in principle
observable in heavy ion collisions, namely an  anticorrelation between the direction
of the CME current for light quarks in a given event and the direction of the 
kick given to the momentum of all the heavy
quarks and antiquarks in that event.
}

\date{\today}

\begin{document}

\maketitle

\section{Introduction and Summary}

The analysis of how a heavy quark moving through the strongly coupled
liquid quark-gluon plasma produced in ultrarelativistic heavy ion collisions
loses energy is motivated by heavy ion collision experiments, in which these
heavy quarks are used as probes of the plasma and, for those
that lose enough energy, as tracers
that follow its flow.
If one assumes that the interactions between the heavy quark
and the plasma are weak, then perturbative methods that were
first introduced in Ref.~\cite{Dokshitzer:2001zm} can be employed
to analyze heavy quark energy loss.
However, the discovery that the plasma produced in heavy ion
collisions is itself a strongly coupled liquid has raised
the question of how to understand the real-time dynamics
of heavy quarks in a strongly coupled non-Abelian plasma.
Treating all aspects of the dynamics as strongly coupled is
of value first as a benchmark and second because it means that rigorous
calculations of novel effects become tractable
in plasmas with a gravitational dual. 

The simplest plasma in which one can calculate the
rate of energy loss of a heavy quark is that in
strongly coupled ${\cal N}=4$ supersymmetric Yang-Mills (SYM)
theory in the large number of colors (large $N_c$) limit, whose
plasma with temperature $T$
is dual to classical gravity in a 4+1-dimensional spacetime
that contains a 3+1-dimensional horizon with Hawking temperature $T$ and
that is asymptotically Anti--de Sitter (AdS) spacetime~\cite{Maldacena:1997re,Witten:1998qj}. 
In the dual gravitational description, the heavy quark is represented by a string 
moving through the AdS black hole spacetime, trailing behind its endpoint that
follows the trajectory of the infinitely heavy quark along the
boundary of the 
AdS~\cite{Karch:2002sh,Herzog:2006gh,Gubser:2006bz,CasalderreySolana:2006rq}.
The earliest work on heavy quark dynamics in the equilibrium plasma of
strongly coupled ${\cal N}=4$ SYM
theory~\cite{Herzog:2006gh,Gubser:2006bz,CasalderreySolana:2006rq} yielded
determinations of the drag force felt by a heavy quark moving through the static plasma
and the diffusion constant that governs the subsequent diffusion of the heavy quark 
once its initial motion relative to the static fluid has been lost due to drag.
This work has been generalized in many directions since then.
We will in particular need the modifications of the spacetime metric that
describe a flowing, hydrodynamic, plasma in which there are gradients of
the fluid properties as a function of space and time~\cite{Bhattacharyya:2008jc,Erdmenger:2008rm,Banerjee:2008th}. 
The corresponding modifications of the drag force were worked out to
leading order in the fluid gradients
in Ref.~\cite{Lekaveckas:2013lha}. However, in this calculation the possibility of
a nonzero density of some fermion species, and a corresponding
nonzero chemical potential, was not taken into account.

We begin in Section~\ref{sec:Fluid} by introducing the dual
gravitational description of a strongly coupled plasma with
both a chemical potential and fluid gradients, working to
leading nontrivial order in both $\mu/T$ and the fluid gradients.
At the same time, we introduce the dual gravitational description
of the axial anomaly, relevant if the chemical potential is
either that for left-handed quarks or for right-handed quarks.
In Section~\ref{sec:DragForce} we turn off the
anomaly, for example as appropriate if $\mu$ is the chemical
potential for baryon number, and calculate the 
corrections to the drag force in powers of $\mu/T$, working
to first order in fluid gradients.  This yields the (straightforward
although laborious) extension of the results of Ref.~\cite{Lekaveckas:2013lha}
to the case of a plasma with nonzero $\mu$.
The results of this Section constitute quantitative
modifications to the drag force, but they do not
introduce qualitatively new effects.

In Section~\ref{sec:Anomalous}
we analyze a chiral plasma. That is, we take $\mu$ to be the chemical
potential for either left-handed or right-handed quarks, and turn on
the anomaly.  In the dual gravitational theory
this means we turn on the Chern-Simons term in the holographic
action that we introduced in Section~\ref{sec:Fluid}.
This term gives rise to novel chiral
contributions (contributions that change sign
if the plasma contains right-handed quarks as opposed to left-handed quarks)
to the charge and entropy currents 
and to the stress-energy 
tensor~\cite{Erdmenger:2008rm,Son:2009tf}, reproducing the
chiral magnetic effect (CME) and chiral vortical effect (CVE)
that had already been introduced  without the
use of holography~\cite{Vilenkin:1980fu,Kharzeev:2007jp,Fukushima:2008xe,Son:2009tf}. 
These anomalous contributions to the hydrodynamic motion
of a chiral fluid have been discussed 
widely~\cite{Erdmenger:2008rm, Son:2009tf,Buividovich:2009wi,Buividovich:2010tn,Kharzeev:2010gr,Neiman:2010zi,Sadofyev:2010is,Kharzeev:2010gd,Amado:2011zx,Landsteiner:2011cp,Kharzeev:2011ds,Hoyos:2011us,Landsteiner:2011iq,Nair:2011mk,Chapman:2012my,Banerjee:2012iz,Jensen:2012jy,Kirilin:2012mw,Eling:2012xa,Stephanov:2012ki,Kalaydzhyan:2012ut,Zakharov:2012vv,Son:2012zy,Chen:2012ca,Akamatsu:2013pjd,Megias:2013joa,Khaidukov:2013sja,Kirilin:2013fqa,Buividovich:2013hza,Avdoshkin:2014gpa}.
These effects originate in topological aspects
of the gauge theory~\cite{Kharzeev:2007jp,Fukushima:2008xe,Son:2009tf,Stephanov:2012ki,Son:2012zy,Chen:2012ca}, as is of course the case for the axial 
anomaly itself~\cite{abj}.

All the previously analyzed
consequences of the axial anomaly in a chiral plasma --- the CME and its cousins --- concern
the motion of light quarks; in
fact the quarks are usually assumed to be massless. We show in Section~\ref{sec:Anomalous}
that there are anomalous 
contributions to the drag force on an infinitely heavy quark that finds itself in a chiral plasma.
The standard CME and related effects involve the generation of anomalous currents
parallel or anti-parallel to an external magnetic field or the angular velocity vector of the fluid~\cite{Vilenkin:1980fu,Kharzeev:2007jp,Fukushima:2008xe,Son:2009tf}. 
The anomalous 
contributions to the drag force that we compute, order by order in $\mu/T$, have the
same feature.  This means that in the presence of a nonzero density of
heavy quarks that are initially at rest in the local fluid rest frame,
the chiral drag force that we compute can yield a new contribution
to the CME electric current, even though the heavy quarks themselves 
do not participate in the CME. 
Note, however, that if the plasma
features equal and opposite number densities of some heavy
quark and its antiquark, or densities of two species of heavy
quarks with opposite electric charges, all the heavy quarks
and antiquarks feel a chiral drag force acting
in the same direction, and no electric current is generated.
For example, the fluid produced in a heavy ion collision is
seeded with equal numbers of charm and anti-charm quarks, and
equal numbers of bottom and anti-bottom quarks. Since in any volume of
the plasma in which there is an excess of, say, right-handed light quarks
all the heavy quarks and antiquarks
feel a chiral drag force in the same direction, the chiral drag force
does not result in an electric current.  Of course, if there were a nonzero
chemical potential for some species of heavy quark, meaning an excess of those heavy quarks
relative to their antiquarks, the push on all heavy quarks and antiquarks from
the chiral drag force would result in a heavy quark contribution to the electric current.
This would be an example of a correction to the CME or CVE currents, as has
been found in other contexts~\cite{Zakharov:2012vv,Akamatsu:2013pjd,Khaidukov:2013sja,Kirilin:2013fqa,Buividovich:2013hza,Avdoshkin:2014gpa}.

It at first seems odd to find a nonvanishing chiral drag force on a heavy quark
even when the heavy quark is at rest in the
local fluid rest frame. 
We show in Section~\ref{sec:Dissipation}
that (as long as we neglect the gravitational anomaly) 
the resolution is related to the previously known fact that in the local
fluid rest frame there are anomalous contributions
to the entropy current,
since if we go instead to a frame in which the
local entropy current vanishes at the location of the
heavy quark we find no chiral drag force
on the heavy quark.  In this frame, the heavy quark
at rest is immersed in a flowing fluid, with nonzero momentum and charge currents that are
both proportional to the axial anomaly coefficient, but the heavy quark feels no
drag force.  If we think of the heavy quark as a defect placed
in these propagating anomaly-induced streams, the fact that the heavy
quark feels no drag force is a direct consequence of the dissipationless
nature of the CME current.
The nondissipative character of the chiral magnetic effect 
has been discussed before~\cite{Kharzeev:2011ds}. 
Our calculation of the chiral drag force provides direct evidence for 
this fundamental attribute of the chiral magnetic effect.


In Section~\ref{sec:Phenomenology} we discuss possible phenomenological consequences
of the chiral drag force.   The basic effect is that  the heavy
quarks and antiquarks in a heavy ion collision in which
there has been a fluctuation resulting in an excess of right-handed (left-handed) light quarks
will feel a force that pushes them in a direction perpendicular to the reaction plane
that is antiparallel (parallel) to the direction of any magnetic field or
fluid angular velocity vector.  We shall show that the effects are small. Furthermore, 
they will
average out in an ensemble of events.
And, mesons containing heavy
quarks are not so numerous in any single heavy ion collision. For all these reasons, it will be difficult
to separate the effects of a small force that acts in the same direction on all 
the heavy quarks and antiquarks in one event from random forces that act differently on different
heavy quarks.  Perhaps clever correlation observables can be found.
One possibility is to utilize the fact that in each event the direction of
the kick that all the heavy quarks and antiquarks in that event receive, regardless
of their electric charge, is opposite to the direction of the CME electric current in the light quark sector.

\section{Holographic fluid}
\label{sec:Fluid}

To compute the drag force on a heavy quark moving through the strongly
coupled plasma of ${\cal N}=4$ SYM theory with a nonzero chemical potential and with spatial and temporal gradients
in the fluid velocity, temperature and chemical potential
we need to start with the perturbations to the dual gravitational
theory that correspond to hydrodynamic flow~\cite{Erdmenger:2008rm,Banerjee:2008th}. 
We shall present the corresponding 4+1-dimensional bulk metric in
this Section, delaying the introduction of the heavy quark that
we are interested in to Section~\ref{sec:DragForce}.
The dual gravitational 
theory is described by the 4+1-dimensional Einstein-Maxwell action 
\begin{eqnarray}
S=-\frac{1}{16\pi G_5}\int d^5 x \sqrt{-G}\left(R+12-\frac{1}{4}F^2\right)
- \frac{\kappa}{48\pi G_5}  \int d^5 x \epsilon^{MNOPQ}A_MF_{NO}F_{PQ}
\end{eqnarray}
whose equations of motion are
\begin{eqnarray}
R_{MN}+4G_{MN}+\frac{1}{2}F_M^{~~K}F_{KN}+\frac{1}{12}G_{MN}F^2=0\notag\\
\partial_N\left(\sqrt{-G}F^{NM}\right)+\kappa\epsilon^{MNOPQ}F_{NO}F_{PQ}=0\ .
\label{eq:EofM}
\end{eqnarray}
Here, $G_5$ is the 4+1-dimensional Newton constant
which according to the holographic dictionary, see e.g.~Ref.~\cite{Erdmenger:2008rm}, is related
to the number of colors in the boundary gauge theory
by $G_5=\frac{\pi}{2 N_c^2}$.
Furthermore, $A_M$ and $F_{MN}$ are a 4+1-dimensional vector potential
and the corresponding field strength and $\kappa$ is the 4+1-dimensional
Chern-Simons coupling, which is dual to the axial anomaly coefficient in
the boundary gauge theory.  
In the case of ${\cal N}=4$ SYM theory
at strong coupling, 
\begin{equation}
\kappa=-\frac{1}{4\sqrt{3}}\ . 
\label{KappaValue}
\end{equation}
(See e.g.~Refs.~\cite{Chamblin:1999tk,Cvetic:1999ne,Son:2009tf}). 
We should mention that we have chosen units in which the AdS radius $R_{\rm AdS}$ has been set to $1$,
meaning that our holographic dictionary coincides with that in Ref.~\cite{Erdmenger:2008rm}.
The equations of motion (\ref{eq:EofM})
have a static black hole solution that corresponds, in the boundary theory,
to a plasma with some nonzero temperature $T$ and some nonzero 
density of right-handed fermions with
chemical potential $\mu$.
Note that $\kappa$ is the bulk Chern-Simons coupling constant and
it gives the boundary theory the axial anomaly.
If we want a model for a theory like QCD in which there are both left- and right-handed
quarks, we need to introduce two bulk vector potentials, one with $\kappa=-\frac{1}{4\sqrt{3}}$
corresponding to $\mu_R$ and the other with $\kappa=+\frac{1}{4\sqrt{3}}$ corresponding
to $\mu_L$.
If we then wish to consider only the case $\mu_L=\mu_R$, we could define
$\mu_V\equiv(\mu_R+\mu_L)/2$, set the two bulk vector potentials equal to each other, and ignore $\kappa$.  
Of course, $\kappa$ becomes relevant in any circumstance in which $\mu_A\equiv (\mu_R-\mu_L)/2 \neq 0$.


In Eddington-Finkelstein coordinates, the metric and bulk gauge field
that describe a static black hole 
of mass $M$ and charge $Q$ take the form
\begin{eqnarray}
\label{g0}
ds^2&=&-r^2f(r)u_\mu u_\nu dx^\mu dx^\nu+r^2P_{\mu\nu}dx^\mu dx^\nu-2u_\mu dx^\mu dr\notag\\
A_\mu&=&-\frac{\sqrt{3}Q}{r^2}u_\mu
\end{eqnarray}
with
\begin{equation}
f(r) \equiv 1-\frac{M}{r^4}+\frac{Q^2}{r^6}\,,
\end{equation}
where 
$u_\mu$ is a constant vector satisfying $u_\mu u^\mu=-1$,
where $P_{\mu\nu}\equiv \eta_{\mu\nu}+u_\mu u_\nu$, 
and where we are working in
axial gauge $A_r=0$.
The energy density, charge density, entropy density, temperature and chemical potential
of the boundary theory strongly coupled plasma can be related to the $M$ and $Q$ of
the dual black hole as follows~\cite{Chamblin:1999tk,Cvetic:1999ne}. 
$\mu$ and $T$ are given by
\begin{equation}
\mu=\frac{\sqrt{3}Q R_{\rm AdS}}{r_+^2}\,, \qquad 
T=\frac{r_+}{2\pi R_{\rm AdS}^2}\left(2-\left(\frac{r_-}{r_+}\right)^2-\left(\frac{r_-}{r_+}\right)^4\right)\,,
\end{equation}
where
$r_+$ and $r_-$  are the larger and smaller real solutions of the equation $f(r)=0$,
and  where we have temporarily restored the factors of $R_{\rm AdS}$.  These relations can
be rewritten as
\begin{equation}
\frac{r_+}{R_{\rm AdS}^2} \equiv \frac{\pi T }{2}\left(1+\sqrt{1+\frac{2\mu^2}{3\pi^2 T^2}}\right) \quad {\rm and} \quad
r_-^2 \equiv \frac{1}{2}r_+^2\left(-1+\sqrt{9-\frac{16}{\left(1+\sqrt{1+\frac{2\mu^2}{3\pi^2 T^2}}\right)}}\right)\, .
\label{rplus}
\end{equation}
The energy density, charge density and entropy density are given in terms of 
$\mu$ and $T$ by~\cite{Erdmenger:2008rm}
\begin{eqnarray}
\label{thermodynamics}
\varepsilon=\frac{3N_c^2}{8\pi^2}  \frac{r_+^3}{R_{\rm AdS}^6}\left(3 \frac{r_+}{R_{\rm AdS}^2}-2\pi T\right) \,,\qquad
\rho=\frac{\mu N_c^2}{4\pi^2} \frac{r_+^2}{R_{\rm AdS}^4}\,,\qquad s=\frac{N_c^2}{2\pi}\frac{r_+^3}{R_{\rm AdS}^6}\,,
\end{eqnarray}
where $N_c$ is the rank of the gauge group and we are working throughout in
the large-$N_c$ limit.   The boundary theory plasma is conformal, meaning
that it has 
$T^\mu_\mu=0$,  and so has an equation of state $\varepsilon=3P$, and has bulk viscosity $\zeta=0$. 
Its shear viscosity can be calculated via gauge/gravity duality and is given by
$\frac{\eta}{s}=\frac{1}{4\pi}$~\cite{Policastro:2001yc,Kovtun:2004de}.

Analogously to the way that hydrodynamics is usually derived,
the way to find a bulk metric that is the dual gravitational description
of a flowing strongly coupled plasma is to look for a solution to
the bulk Einstein-Maxwell equations in which 
$T$, $\mu$ and $u^\mu$ are all slowly varying functions of space
and time, and to organize
the calculation via a gradient expansion.
The metric is expanded in powers of boundary gradients 
\begin{equation}
G_{\mu\nu}=G^{(0)}_{\mu\nu}+G^{(1)}_{\mu\nu}+{\cal O}(\partial^2)\,,
\end{equation}
where $G^{(0)}_{\mu\nu}$ is defined in (\ref{g0}) and where 
$G_{\mu\nu}^{(1)}$ contains all possible gradient structures of first order, 
with unknown coefficient functions. 
The solution has been obtained up to first order in gradients and is given by~\cite{Erdmenger:2008rm}
\begin{eqnarray}
\label{metric}
ds^2&=&-r^2f(r)u_\mu u_\nu dx^\mu dx^\nu+r^2P_{\mu\nu}dx^\mu dx^\nu-2u_\mu dx^\mu dr+r^2F(r)\sigma_{\mu\nu}dx^\mu dx^\nu\notag\\&~&\qquad+r^2j_\sigma\left(P^\sigma_\mu u_\nu+P^\sigma_\nu u_\mu\right)dx^\mu dx^\nu+\frac{2}{3}r(\partial\cdot u) u_\mu u_\nu dx^\mu dx^\nu
\end{eqnarray}
where
\begin{equation}
\sigma^{\mu\nu}\equiv  P^{\mu\alpha}P^{\nu\beta} \left( \partial_\alpha u_\beta +\partial_\beta u_\alpha \right)
-\frac{2}{3}P^{\mu\nu}\partial\cdot u
\end{equation}
and
\begin{equation}
\label{eq:jsigma}
j_\sigma\equiv -\frac{1}{r}(u\cdot\partial) u_\sigma+\frac{2\sqrt{3}Q^3\kappa}{M r^6}\ell_\sigma+J(r)\partial_\sigma\frac{\mu}{T}\,,
\end{equation}
where $\ell_\mu\equiv \epsilon_{\mu\nu\alpha\beta}u^\nu\partial^\alpha u^\beta$
is the four-vector containing the vorticity of the fluid, 
and where the functions $F(r)$ and $J(r)$ are, up to leading nontrivial order in $\mu$, given by
\begin{eqnarray}\label{Fdefn}
F(r)&\equiv&\frac{1}{4\pi T}\left[2\arctan{\left(\frac{\pi T}{r}\right)-\log\left(\frac{r^4}{(r+\pi T)^2(r^2+\pi^2 T^2)}\right)}\right]\notag\\
&~&\quad+\frac{\mu^2}{24\pi^2 T^2}\Biggl[-\frac{3}{T}+\frac{4\pi T}{r^2}+\frac{2}{r+\pi T}+\frac{4(r+\pi T)}{r^2+\pi^2 T^2}\notag\\
&~&\quad\qquad+\frac{6}{\pi T}\left[\arctan\left(\frac{r}{\pi T}\right) + \log\left(\frac{ r^2}{r^2+\pi^2 T^2}\right)\right]\Biggr]+{\cal O}\left(\mu^4\right)
\end{eqnarray}
and
\begin{eqnarray}
J(r)&\equiv&\frac{\mu}{24\pi^3 r^4 T^2}\Biggl[2\pi^2 T^2 r(3r-2\pi T)-3(r^4-\pi^4 T^4)\left[2\arctan\left(\frac{\pi T}{r}\right)-\log\left(\frac{(r+\pi T)^2}{r^2+\pi^2 T^2}\right)\right]\Biggr]\notag\\
&~&\qquad\quad+{\cal O}\left(\mu^3\right)\,.
\end{eqnarray}
Note that since $Q\propto\mu$ the anomalous term in $j_\sigma$ that is proportional to $\kappa\ell_\sigma$ is
of order $\mu^3$, while the leading $\mu$-dependence in the metric (\ref{metric}) coming
from $F(r)$ and $J(r)$ is of order $\mu^2$.  We will see the effects of the terms of order $\mu^2$
in Section~\ref{sec:DragForce} and shall introduce the effects of order $\mu^3$ that originate
from the axial anomaly in Section~\ref{sec:Anomalous}.  In Section~\ref{sec:Anomalous} we
shall also introduce effects that originate from the axial anomaly 
that are proportional to an external magnetic field, rather than to the fluid vorticity.
Doing so will require adding an additional term in $j_\sigma$, a term that is of order
$\mu^2$, and that is proportional to $\kappa$ and the magnetic field.
Note that, as in Ref.~\cite{Erdmenger:2008rm}, we are using the Landau definition of
the fluid velocity $u^\mu$, such that the Lorentz frame in which
$u^\mu=(1,\vec{0})$ corresponds to the frame in which
the fluid momentum $T^{0i}$ vanishes.  In this local fluid rest frame,
there can still be nonzero charge currents or entropy currents, as we shall
see later.

To the same order in $\mu/T$ and gradients as above, the bulk gauge field now takes the form
\begin{eqnarray}
A_\mu=-\frac{\sqrt{3}Q}{r^2}u_\mu+\frac{6\kappa Q^2}{M r^2}\ell_\sigma+a(r)\partial_\sigma\frac{\mu}{T}
\end{eqnarray}
where we are still in $A_r=0$ gauge and 
where $a(r)$ is a known function that can be found in Ref.~\cite{Erdmenger:2008rm}.  However, the bulk gauge field
does not affect the string that we will introduce
in the next Section that constitutes the dual description of the heavy quark,
and for this reason we do not require $A_\mu$ for our considerations.


In most contexts relevant to heavy ion collisions, $\mu/T$ is small.  If $\mu$ represents 
an excess of
$\mu_R$ over $\mu_L$ or vice versa, these only arise due to fluctuations or anomalous effects and are
certainly small compared to $T$.  If, as in Section~\ref{sec:DragForce}, we treat a nonchiral
plasma and think of $\mu$  as representing the chemical potential for quark number (which
in turn is one-third
that for baryon number), $\mu/T$ can be as large as $\sim 1$ in the lowest energy heavy
ion collisions possible at RHIC but in higher energy collisions it is substantially
smaller than 1. We will see in Section~\ref{sec:Phenomenology} that
the effects of interest are suppressed by powers of $\mu_A/(\pi T)$ or $\mu_V/(\pi T)$;
even if the quark number chemical potential 
$\mu_V$ can get as large as $\sim T$ it is always small compared to $\pi T$.
We will therefore work perturbatively in powers of $\mu/T$.

\section{Drag Force}
\label{sec:DragForce}

In this Section we shall only consider $\mu_L=\mu_R$ and shall therefore set $\kappa=0$ and think of $\mu$ as representing the quark
number chemical potential $(\mu_L+\mu_R)/2$. 

The drag force has been calculated in a static
plasma with $\mu=0$ and no gradients in Refs.~\cite{Herzog:2006gh,Gubser:2006bz,CasalderreySolana:2006rq}.
The basic picture
of heavy quark dynamics that emerges, with all but the initially most energetic heavy quarks
being rapidly slowed by drag and then becoming tracers diffusing within the (moving) fluid, is qualitatively
consistent with early experimental investigations~\cite{Adare:2006nq}.  For a 
review, see Ref.~\cite{CasalderreySolana:2011us}.
Subsequently, the holographic calculational techniques were generalized to any static plasmas
whose gravitational dual has a 4+1-dimensional metric that depends only on the
holographic (i.e.~`radial') coordinate in Ref.~\cite{Herzog:2006se} and heavy quark energy loss
and diffusion has by now been investigated in the equilibrium plasmas of many gauge theories
with gravitational duals~\cite{Caceres:2006dj, Caceres:2006as, Matsuo:2006ws, Nakano:2006js, Talavera:2006tj, Gubser:2006qh, Bertoldi:2007sf, Liu:2008tz,Gursoy:2009kk, HoyosBadajoz:2009pv, Bigazzi:2009bk, NataAtmaja:2010hd, Chernicoff:2012iq, Fadafan:2012qu, Giataganas:2012zy}. 
In particular, the drag force on a heavy quark in a static
plasma with $\mu\neq 0$ was calculated in Refs.~\cite{Herzog:2006se,Caceres:2006dj}.
We reproduce these results in Section~\ref{sec:Rest}.  More recently,
the drag force
on a heavy quark moving through the far-from-equilibrium
matter present just after the collision of two sheets of energy
density~\cite{Chesler:2010bi} has been calculated and
compared to that in static strongly coupled plasma in
equilibrium~\cite{Chesler:2013cqa}. 
Motivated initially by the need to understand these results,
in Ref.~\cite{Lekaveckas:2013lha}
the drag force was computed 
in a fluid with $\mu=0$ in which there are spatial and temporal
gradients in the fluid temperature and flow velocity, to
leading order in these fluid gradients. In Section~\ref{sec:Gradients},
we shall reproduce these results and shall
compute the leading
order effects of fluid gradients in a strongly coupled ${\cal N}=4$ SYM plasma with $\mu\neq 0$.
(As an aside, we note that the effects of fluid gradients on photon emission
have also been investigated~\cite{Mamo:2014ema}.)


\subsection{Fluid at rest}
\label{sec:Rest}

According to Refs.~\cite{Herzog:2006gh,Gubser:2006bz}, 
the dual picture of an (infinitely) heavy quark or antiquark moving through the strongly coupled plasma 
of ${\cal N}=4$ SYM theory corresponds to a string whose endpoint moves on the boundary of AdS along 
the trajectory of the heavy quark or antiquark of interest.  The string trails ``downward'' into
the bulk spacetime, toward the black hole horizon.  It also trails behind the
moving quark or antiquark.  The drag force is defined as the force needed to move
the heavy quark with constant velocity $v$ through the plasma, so we take the
endpoint of the string to move along the AdS boundary with constant velocity $v$.
The drag force is determined in the dual gravitational
description by the shape of the trailing string.  Because the string
trailing behind a heavy quark or a heavy antiquark moving with the same velocity
through the same fluid has the same shape, they feel the
same force.  This conclusion applies throughout: the introduction
of gradients in the fluid, magnetic fields, and anomalous effects later in this
Section and in Section 4 does not change the fact that heavy quarks and antiquarks
feel the same force.  We shall therefore henceforth refer only to heavy quarks.

Without loss of generality one can consider the heavy quark velocity $\vec v$ to be directed along the 
$x$-direction and because we are considering a static plasma it makes sense to work in 
the rest frame of the medium, generalizing this only later. 
In this setup the probe string is described by the Nambu-Goto action
\begin{eqnarray}
S_{NG}=-\frac{\sqrt\lambda}{2\pi}\int d\tau d\sigma\sqrt{-g(\tau,\sigma)},
\end{eqnarray}
where $\lambda$ is the 't Hooft coupling, which we are assuming is large, and with 
$g(\tau,\sigma)$ the determinant of the induced worldsheet 
metric. The worldsheet metric is defined as 
$g_{ab}\equiv G_{AB}\partial_a x^A\partial_b x^B$ with $x^A(\tau,\sigma)$ the string profile. Here, upper case Latin indices correspond to 4+1-dimensional spacetime  indices and lower case indices correspond
 to the 1+1-dimensional worldsheet metric indices. 
 We also have freedom in the choice of the parametrization of the worldsheet and it is convenient to 
 take $$t(\tau,\sigma)=\tau\,, \qquad r(\tau,\sigma)=\sigma\,.$$ 
 For a quark whose trajectory in the boundary theory
 is $x=vt$, the ansatz for the shape of the string in the bulk 
 that corresponds to the word ``trailing'' 
 is
 \begin{equation}
 x(\tau,\sigma)=v\tau+\xi(\sigma),
 \end{equation}
 with $y(\tau,\sigma)=z(\tau,\sigma)=0$.

The string profile $\xi(\sigma)$ is found 
by extremizing the Nambu-Goto action and solving the resulting Euler-Lagrange equation 
\begin{eqnarray}
\partial_\tau\left(\frac{\delta {\cal L}}{\delta\partial_\tau \vec x}\right)+\partial_\sigma\left(\frac{\delta {\cal L}}{\delta\partial_\sigma \vec x}\right)=\left(\frac{\delta {\cal L}}{\delta \vec x}\right)\ .
\label{eq:EulerLagrange}
\end{eqnarray}
We shall solve this 
for the string profile 
order-by-order in powers of $\mu/T$. 
The $\mu=0$ (or equivalently zero quark number) solution was obtained in Refs.~\cite{Herzog:2006gh,Gubser:2006bz}. 
We can write the expansion about the $\mu=0$ solution in Eddington-Finkelstein coordinates as
\begin{eqnarray}
\vec{x}^{0,2}=\vec v \,\tau+\vec{v}\,\frac{ \pi-2\arctan \left(\frac{\sigma}{\pi T}\right)}{2\pi T}+\vec\xi^{0,2}(\sigma)
+{\cal O}\left(\mu^4\right)\ ,
\end{eqnarray}
where the $0,2$ superscript refers to zeroth order in gradients and second order in $\mu/T$
and where the direction of $\vec\xi^{0,2}$ is the same as the direction of $\vec v$, which
we are taking to be the $x$ direction.  From the Euler-Lagrange equation (\ref{eq:EulerLagrange}),
we find that the equation of motion for $\xi_x^{0,2}(\sigma)$ takes the form
\begin{eqnarray}
\partial_\sigma(\sigma^4-\pi^4T^4\gamma^2)\partial_\sigma \xi^{0,2}_x(\sigma)= {\rm RHS}
\end{eqnarray}
where the left-hand side that we have given explicitly has the same form also at higher orders
in $\mu/T$ while the right-hand side, an expression involving no derivatives with respect to $\sigma$,
is different at each order. Since the string endpoint is constrained to move along the constant velocity
trajectory $x=vt$,
we require $\xi^{0,2}_x(\infty)=0$. 
Upon solving the equation of motion,  we find that the string profile
to order $\mu^2/T^2$ and to zeroth order in gradients takes the form
\begin{eqnarray}
\vec{x}&=&\vec v \tau+\frac{\vec{v}(\pi-2\arctan\left(\frac{\sigma}{\pi T}\right))}{2\pi T}\notag\\
&~&\qquad
+\frac{\vec{v}}{12\pi^3 T}\left(\frac{\mu}{T}\right)^2 \Biggl[\pi\left(-3-\frac{1}{\gamma^{5/2}}+\frac{2T(1+\gamma^2)}{\gamma^2\sigma}+\frac{4 T \sigma}{\pi^2 T^2+\sigma^2}\right)\notag\\
&~&\qquad\qquad\qquad\qquad\qquad\qquad+6\arctan\left(\frac{\sigma}{\pi T}\right)+\frac{2}{\gamma^2\sqrt{\gamma}}\arctan\left(\frac{\sigma}{\pi T\sqrt{\gamma}}\right)\Biggr]
\label{eq:ProfileNoGradient}
\end{eqnarray}
where $\gamma=1/\sqrt{1-v^2}$.
 In the zero chemical potential limit, (\ref{eq:ProfileNoGradient}) is the solution obtained in Refs.~\cite{Herzog:2006gh,Gubser:2006bz}. 
In solving the equations of motion to obtain (\ref{eq:ProfileNoGradient}),
we fixed the integration constants by requiring that the worldsheet must be regular 
at the worldsheet horizon, to zeroth order in $\mu$ located at 
$\sigma=\pi T\gamma^\frac{1}{2}$, and by requiring that the string endpoint follows the desired trajectory,
as already mentioned.

With the string profile in hand, we can now compute the drag force.
The drag force is defined as (see e.g. Refs.~\cite{Herzog:2006gh,Gubser:2006bz,Lekaveckas:2013lha})
\begin{eqnarray}
f^\mu(\tau) \equiv -\lim_{\sigma\to\infty}\eta^{\mu\nu}\Pi^\sigma_\mu(\tau,\sigma)
\end{eqnarray}
 where the string momentum is 
 \begin{equation}
 \Pi^\sigma_\mu \equiv -\frac{\sqrt{\lambda}}{2\pi}G_{\mu N}\frac{1}{\sqrt{-g}}\left[g_{\tau\sigma}\partial_\tau X^N-g_{\tau\tau}\partial_\sigma X^N\right]
 \end{equation} 
 and we are interested 
 in the instantaneous drag force at $\tau=0$. (Note that $f^\mu$ is not a Lorentz 4-vector but one could introduce the proper force 
 $\gamma f^\mu$ which has appropriate transformation properties.) 
 Upon substituting the string profile (\ref{eq:ProfileNoGradient}) into the definition of the string 
 momentum we have
\begin{eqnarray}
\lim_{\sigma\to\infty} \Pi^\sigma_\mu(0,\sigma)=-\frac{T^2\sqrt{\lambda}}{12\pi}  \left[6\pi^2\gamma-(1-3\gamma)\left(\frac{\mu}{T}\right)^2\right] \left( 1-\frac{1}{\gamma^2} ,\vec{v}\right)
\end{eqnarray}
from which we find that the drag force is given by 
\begin{eqnarray}
f^\mu_{(0)}=\frac{T^2 \sqrt{\lambda}}{12\pi}\frac{1}{\gamma^2}(\gamma w^\mu-\delta^\mu_0)\left(6\pi^2\gamma-(1-3\gamma)\left(\frac{\mu}{T}\right)^2\right)
\label{OrdinaryDrag}
\end{eqnarray}
where $w^\mu=\gamma(1,\vec v)$ is the four velocity of the quark.   This result was first obtained 
in Ref.~\cite{Herzog:2006se}.
Note that this force,
conventionally called the drag force, is defined as the force
that an external agent must exert on the heavy quark in order to
keep it moving at constant velocity $\vec v$.  We see from the result (\ref{OrdinaryDrag}) that
$\vec f$ points in the same direction as $\vec v$.  The force that the fluid
itself exerts on the heavy quark is $-\vec f$, in the $-\vec v$ direction.

We have obtained the drag force in the fluid rest frame, where $u^\mu=(1,\vec 0)$.  We can
now Lorentz-transform $\gamma f^\mu$ to a general frame, and upon doing so we find
\begin{eqnarray}
f^\mu_{(0)}=-\frac{\sqrt{\lambda}}{2\pi}\frac{\pi^2 T^2}{\gamma}(s w^\mu+u^\mu)\left(1+\frac{(1+3s)}{6s}\left(\frac{\mu}{\pi T}\right)^2\right)
\end{eqnarray}
where $s\equiv u\cdot w$ is 
the only Lorentz scalar that can arise at zeroth order in gradients.
It also can be shown by direct calculation,
starting from the definitions of the drag force and the string momentum, that $w_\mu f^\mu(\tau)=0$.
Note that in the nonrelativistic limit $\gamma\rightarrow 1$ we have $s=-1$ and
the $\mu$-dependence of the drag force is $\sim \left(1+\frac{1}{3}(\frac{\mu}{\pi T})^2\right)$
whereas in the ultra-relativistic limit $\gamma\rightarrow\infty$ we have $s=-\infty$
and 
the $\mu$-dependence of the drag force is $\sim \left(1+\frac{1}{2}(\frac{\mu}{\pi T})^2\right)$.
This means that in the presence of a nonzero chemical potential, the force
required to move the heavy
quark through the strongly coupled plasma
is strictly speaking no longer a drag force, since the magnitude of the
force is no longer proportional to the momentum of the heavy quark.
This is also the case in the presence of nonzero fluid gradients, as shown
in Ref.~\cite{Lekaveckas:2013lha}.  By convention, we will nevertheless continue to refer
to the force needed to move the heavy quark through the fluid as a drag force.
As a side remark here, but a side remark that we will need in Eq.~(\ref{OrdinaryDragGeneral}) in Section 5, 
note from (\ref{rplus}) that $r_+\sim \pi T  \left(1+\frac{1}{6}(\frac{\mu}{\pi T})^2\right)$,
indicating that in the nonrelativistic limit the drag force is proportional to $r_+^2$.


\subsection{Corrections to the Drag Force Due to Fluid Gradients} 
\label{sec:Gradients}

The drag force was calculated to first order in fluid gradients at zero $\mu$  in 
Ref.~\cite{Lekaveckas:2013lha}. 
We can follow the same logic.  We must first obtain the string profile to first order in 
fluid gradients.
Since the string profile at zeroth order in gradients that we obtained in (\ref{eq:ProfileNoGradient}) 
above contains $\tau$ only at  linear order 
one can make the following ansatz, which satisfies the equation of motion:
\begin{eqnarray}
\vec x^{1,n}=\vec{x}^{0,n}+\tau h^{1,n}(\sigma)+\vec{\xi}^{1,n}(\sigma)\,.
\end{eqnarray}
Here, the expression for the string profile to zeroth order in fluid gradients 
in a general frame can easily be obtained from our previous results and is given by
\begin{equation}
\vec{x}^{0,0}(\tau,\sigma)=\vec{v}\tau-\frac{1}{\pi T}\left(u^0\vec{v}-\vec u\right)\left(\arctan\left(\frac{\sigma}{\pi T}\right)-\frac{\pi}{2}\right)\,.
\end{equation}
$\vec x^{0,1}$ vanishes, and $\vec x^{0,2}$ can also easily be obtained from our previous results. 
It can then  be shown that choosing $h^{1,n}(\sigma)=D_t\vec{x}^{0,n}|_{\tau=0}$, with 
the medium derivative defined as $D_t\equiv\partial_t+v^i\partial_i$, 
cancels 
the $\tau$-dependence in the equation of motion. 
Next, we expand about the zeroth-order-in-gradients solution, see the $\tau$-dependent terms cancel against
contributions introduced by $h^{1,n}(\sigma)$, and find that $\vec{\xi}^{1,n}$ satisfies an
equation of the form
\begin{eqnarray}
\label{eom1}
\partial_\sigma(\sigma^4-\pi^4T^4\gamma^2)\partial_\sigma \xi^{1,n}_i(\sigma)= {\rm RHS} 
\end{eqnarray}
where the right-hand side depends only on $\sigma$, not on derivatives of $\sigma$, and 
where the right-hand side is the same for $i=y$ and $i=z$ but is different for $i=x$.  We shall
not provide the expressions for the right-hand side here as they are lengthy.
%
%
In solving the equations of motion (\ref{eom1}), the boundary conditions are the same
as at zeroth order and again we require regularity at the worldsheet horizon. 
%
%
The full solution for  $\vec{\xi}^{1,0}(\sigma)$ can be found in Ref.~\cite{Lekaveckas:2013lha}.  We have obtained
$\vec{\xi}^{1,2}(\sigma)$, but again these expressions are very lengthy and we will not give them
here, focussing instead on presenting results for the drag force.


To first order in gradients and zeroth order in $\mu$, we reproduce the gradient corrections to
the drag force on the heavy quark that were first obtained in Ref.~\cite{Lekaveckas:2013lha} and that take the form
\begin{eqnarray}
f_{(1,0)}^\mu&=&-\frac{\sqrt{\lambda}}{2\pi}\frac{\pi T}{\gamma}\Biggl[ c_1(s)\Bigl(u^\mu(w\cdot\partial)s-s\partial^\mu s-s(s u^\alpha+w^\alpha)\partial_\alpha U^\mu\Bigr)\nonumber\\
&~&\qquad\qquad\qquad\qquad+c_2(s)U^\mu(\partial\cdot u)-\sqrt{-s}(u\cdot\partial)U^\mu \Biggr]
\label{eq:MindaResult}
\end{eqnarray}
where $s\equiv u^\alpha w_\alpha$ as before and where we have defined
\begin{eqnarray}
c_1(s)&\equiv&\pi/2-\arctan\left( \sqrt{-s}\right)-\pi T F^{0}(s)\notag\\
c_2(s)&\equiv&\frac{1}{3}(\sqrt{-s}+(1+s^2)c_1(s))\notag
\end{eqnarray}
and introduced the projector $U_\mu\equiv u_\mu+s w_\mu$.  Here, the function $F^0(s)$ is obtained
by setting $\mu=0$ in the function $F$ defined in (\ref{Fdefn}). 


Turning now to the corrections that are leading nonzero order in both gradients and $\mu/T$, following the logic
set out above we expand the equations of motion for the string profile in a double expansion in gradients
and $\mu/T$, solve them, and obtain the force from the string profile as we did at zeroth order.  We find
no contribution that is first order in $\mu/T$.  To second order in $\mu/T$ and first order in gradients,
the correction to the drag force is given by
\begin{eqnarray}
f_{(1,2)}^\mu&=&-\frac{\sqrt{\lambda}}{48\gamma\pi^2 }\frac{\mu^2}{T} \times \Biggl[
-c_3(s)\Bigl(u^\mu (w\partial)s-s\partial^\mu s\Bigr) - U^\mu c_4(s)(\partial \cdot u)-\frac{2c_5(s)}{s}\left((w\cdot\partial)\log\frac{\mu}{T}\right) u^\mu
 \notag\\
&~&\quad\qquad\qquad+ U^\mu \left(\frac{16-4\pi}{(-s)^\frac{3}{2}}+2\frac{c_5(s)}{s}\right)(w\cdot\partial)\log\frac{\mu}{T} +U^\mu\left(\frac{4c_1(s)}{s}+\frac{8-2\pi}{(-s)^\frac{5}{2}}\right)(s u^\alpha+w^\alpha)\partial_\alpha s\notag\\
&~&\quad\qquad\qquad+\frac{10-3\pi}{(-s)^\frac{5}{2}}U^\mu(w\cdot\partial)s+\left(s^2c_3(s)+\frac{6s^3+(3\pi-10)s^2+4-\pi}{(-s)^\frac{5}{2}}\right)(u\cdot\partial)U^\mu\notag\\
&~&\quad\qquad\qquad+\left(s c_3(s)-\frac{4(\pi-4)}{(-s)^\frac{3}{2}}\right)(w\cdot\partial)U^\mu+2c_5(s)\partial^\mu\log\frac{\mu}{T}\Biggr]
\end{eqnarray}
where we have defined
\begin{eqnarray}
c_3(s)&\equiv&\frac{1}{(-s)^{\frac{5}{2}}} \Biggl[ 4-\pi+6(-s)^{\frac{3}{2}}-6s^2+4(-s)^{\frac{3}{2}}c_1(s)+3(-s)^{\frac{5}{2}}\log\left(1-\frac{2\sqrt{-s}}{-1+s}\right) \Biggr] \notag\\
c_4(s)&\equiv&\frac{1}{3(-s)^{\frac{3}{2}}}\Biggl[2\Bigl(3\sqrt{-s}+s\bigl(2+3s(-1+\sqrt{-s}+s)\bigr)\Bigr)\notag\\
&~&\qquad\qquad+\sqrt{-s}\left(8c_1(s)-3s(1+s^2)\log\left(1-\frac{2\sqrt{-s}}{-1+s}\right)\right)\Biggr]\\
c_5(s)&\equiv&\frac{1}{\sqrt{-s}} \Biggl[ 4-3s(\sqrt{-s}+2s)+6\sqrt{-s}(s^2-1)\left(c_1(s)+\log\left(1+\frac{1}{\sqrt{-s}}\right)\right)\Biggr]\notag
\end{eqnarray}
To a large extent, all of these contributions to the drag force can be considered corrections to
the $\mu=0$ results (\ref{eq:MindaResult}) from Ref.~\cite{Lekaveckas:2013lha}; they introduce quantitative changes to
the gradient corrections to the drag force, but do not change the story in qualitative ways
beyond introducing $\partial_\mu (\mu/T)$ terms that describe the contributions to
the drag force on the heavy quark due to gradients
in the chemical potential, and the corresponding charge currents.


\section{Anomalous Contributions}
\label{sec:Anomalous}

In the previous Section, we have computed the complete correction to the drag force
to first order in fluid gradients up to order $\mu^2$ in the absence of $\kappa$,  
which is to say in the absence of any chiral anomaly,
as for example for the case where $\mu_L=\mu_R$ and
the $\mu$ in the previous Section represents
$(\mu_L+\mu_R)/2$. In this setting, the next order contributions to the drag force would
either be of order $\mu^4$ or second order in gradients, and as far as we can
see would not introduce any qualitatively new effects.

We now allow $\mu_R$ and $\mu_L$ to differ and hence we turn on  
$\kappa$, the Chern-Simons coupling in the bulk gravitational
theory that describes the axial anomaly in the
boundary gauge theory.   Upon doing so, we shall find contributions
to the drag force on a heavy quark that are qualitatively new, arising
in two ways.
First, we find new effects that arise at order $\mu^3$ because the contribution to the metric (\ref{metric})
proportional to $\kappa$ is proportional to $\mu^3$.
These effects are proportional to $\ell_\mu=\epsilon_{\mu\nu\alpha\beta}u^\mu\partial^\alpha u^\beta$
and hence are first order in gradients.
And, they only arise when $\mu$ represents a chiral chemical potential
like $\mu_L$ or $\mu_R$, meaning that the effects
we shall consider in this Section should  be 
smaller in magnitude than those in the previous Section. 
Second, we find new effects that arise at order $\mu^2$ and are proportional
to $\kappa$ that arise only in the presence of a magnetic field.

We shall also describe the contributions to the drag force that 
are introduced when we include the gravitational anomaly in the
boundary theory.

As is by now well studied~\cite{Erdmenger:2008rm,Son:2009tf,Neiman:2010zi,Sadofyev:2010is,Zakharov:2012vv}, introducing the Chern-Simons coupling in the
bulk corresponds to introducing anomalous contributions to the hydrodynamic
equations for a chiral plasma (a plasma with $\mu_L\neq\mu_R$)
in the boundary theory, contributions that arise because of the axial anomaly
in the gauge theory~\cite{abj}.
The corresponding anomalous transport phenomena have been
studied at weak coupling~\cite{Vilenkin:1980fu,Fukushima:2008xe, Landsteiner:2011cp} as well as at strong 
coupling~\cite{Erdmenger:2008rm,Son:2009tf,Neiman:2010zi,Sadofyev:2010is,Landsteiner:2011cp,Jensen:2012jy,Zakharov:2012vv,Buividovich:2014dha}. 
These anomalous effects can be summarized by noting that in a chiral fluid in the presence of an
external magnetic field $\vec B$ or of nonzero fluid angular
momentum $\vec \Omega$ the axial anomaly causes both vector and
axial currents to flow~\cite{Vilenkin:1980fu,Son:2009tf,Sadofyev:2010is,Landsteiner:2011cp}: 
\begin{eqnarray}
\label{ce}
\vec J_V(x)&=&\frac{\mu_A}{2\pi^2}\vec B+\frac{\mu_V\mu_A}{\pi^2}\vec\Omega\notag\\
\vec J_A(x)&=&\frac{\mu_V}{2\pi^2}\vec B+\left(\frac{\mu_V^2+\mu_A^2}{2\pi^2}+\frac{T^2}{6}\right)\vec\Omega\ ,
\end{eqnarray}
where $\mu_V\equiv (\mu_R+\mu_L)/2$ and $\mu_A\equiv (\mu_R-\mu_L)/2$ are the 
vector and axial chemical potentials. The first and second terms in the vector current have
been named the chiral magnetic effect (CME) and the chiral vortical effect (CVE), respectively.
The expressions
(\ref{ce}) are valid for a $U(1)$ gauge theory; in Section 5 we will generalize them
as appropriate for ${\cal N}=4$~SYM theory.
We note also that in (\ref{ce}) we have left out terms that are higher order in chemical
potentials (order $\mu^2$ in the CME and order $\mu^3$ in the CVE) that we will need, and introduce, later in (\ref{FullChargeCurrent}).
The contribution to $\vec J_A$ that is proportional to $T^2$ arises 
due to the gravitational anomaly (see e.g.~Refs.~\cite{Landsteiner:2011cp, Golkar:2012kb, Jensen:2012kj}).


We shall introduce the four-vector
$B_\mu\equiv\frac{1}{2}\epsilon_{\mu\nu\alpha\beta} u^\nu F^{\alpha\beta}$.  Note
that in the limit in which the fluid velocity is nonrelativistic, $B_\mu$ and $\ell_\mu$ 
take the form $B_\mu=(0,\vec{B})$ and $\ell_\mu=(0, 2\vec \Omega)$ in
the local fluid rest frame, with $\vec{B}$ and $\vec\Omega$ being the (externally applied) magnetic
field and the local angular velocity of the fluid.


\subsection{Chiral Vortical Drag Force}

In this Section, we shall compute the anomalous contributions
to the drag force required to pull a heavy quark through a chiral plasma,
in which currents like (\ref{ce}) are flowing.
We begin by setting the external magnetic field $\vec B=0$, considering
only the effects of vorticity in the chiral fluid.
We derive and solve equations of the form (\ref{eom1}) 
and find that at order $\mu^3$ the string profile contains no term proportional to $\tau$ and
confirm that a contribution at order $\mu^3$ can only come from the presence of the Chern-Simons term
in the bulk metric. Direct calculation then yields a contribution to the string profile given by
\begin{eqnarray}
\vec{\xi}\,^\ell_{1,3}(\sigma)=- \frac{2\kappa T}{3\pi^2\gamma}\left(\frac{\mu}{T}\right)^3\frac{ \pi^4 T^4\gamma^2+\pi^2 T^2\gamma^2(1+\gamma)\sigma^2+\sigma^4 }{\sigma^4(\pi^2 T^2\gamma+\sigma^2)(\pi^2 T^2+\sigma^2)}\,\vec{\ell} 
\end{eqnarray}
and a vorticity-induced contribution to the drag force in a generic frame given by
\begin{eqnarray}
\label{fO}
(f^{\ell})_{1,3}^\mu=\frac{\kappa\sqrt{\lambda} }{3\gamma\pi^3 }  \frac{\mu^3}{T^2}\frac{\ell^\mu+(\ell\cdot w)w^\mu}{s}\,.
\end{eqnarray}
This chiral contribution is directly proportional to the anomalous coefficient $\kappa$, as expected.
In the calculation above, $\mu$ represents $\mu_R$.  We now repeat the calculation for $\mu_L$,
flipping the sign of $\kappa$, and find that when $\mu_L$ and $\mu_R$
are both nonzero the result is obtained by replacing $\mu^3$ in (\ref{fO}) by
$\mu_R^3-\mu_L^3=6\mu_A\mu_V^2 + 2 \mu_A^3$.

\subsection{Chiral Magnetic Drag Force}

Next, we are interested in the anomalous contribution to the drag force due
to the presence of a magnetic field. (We will not attempt the analysis for
the more generic case of background electric and magnetic fields.)
First, we need to modify the gravitational metric in a way that corresponds
to introducing an external magnetic field in the boundary gauge theory, see e.g.~Ref.~\cite{Megias:2013joa}.
This is done by 
adding a term to the static metric (\ref{g0}) that corresponds
to adding a new term to $j_\sigma$ in (\ref{eq:jsigma}) given by
\begin{eqnarray}
j^B_\sigma \equiv \frac{\kappa}{\pi^2 T^2}\left(\frac{\mu}{\pi T}\right)^2C_B(r)B_\sigma,
\label{eq:jBsigma}
\end{eqnarray}
where 
\begin{equation}
C_B(r)\equiv
\frac{2\pi^2 T^2}{r^2}-\frac{\pi^4 T^4}{r^4}+2 \left(1-\frac{\pi^4 T^4}{r^4}\right)\log\left(\frac{r^2}{r^2+\pi^2 T^2}\right)\ .
\end{equation}
The full consideration of hydrodynamics with external electric and magnetic fields can be found in Ref.~\cite{Megias:2013joa}. 
We shall follow the same procedure as above. And, as we are interested only in the effects
due to the presence of the magnetic field we can work to zeroth order in fluid gradients.
Since there is no contribution in it that is linear in $\mu/T$ the string equation of motion (\ref{eom1}) 
simplifies as in the vortical case above and we find a correction to the string profile given by
\begin{eqnarray}
\vec{\xi}\,_{1,2}^{B}(\sigma)=-\frac{\kappa}{\pi^4 T^2}\left(\frac{\mu}{T}\right)^2\frac{\sigma^2(\pi^2 T^2\gamma^2+\sigma^2)C_B(\sigma)-\pi^2 T^2\gamma^2(\pi^2 T^2+\sigma^2)C_B(\pi T\sqrt{\gamma})}{(\pi^4 T^4\gamma^2-\sigma^4)(\pi^2 T^2+\sigma^2)}\vec{B}\,,  
\end{eqnarray}
where we are considering the magnetic field itself to be first order in gradients.  This
results in a magnetic-field-induced
correction to the drag force given by
\begin{eqnarray}
\label{fB}
(f^{B})_{1,2}^\mu=-\frac{\kappa\sqrt{\lambda}}{2\pi^3\gamma} \left(\frac{\mu}{T}\right)^2  s^2~C_B(\pi T\sqrt{-s}) \left(B_\mu+(B\cdot w)w_\mu\right)\,.
\end{eqnarray}
We have excluded the term  
corresponding to the Lorentz force on the quark, as this does not depend on the medium or
on the chiral anomaly and is not of interest to us here.

We then repeat the calculation for $\mu_L$, flipping the sign of $\kappa$, and
hence replace $\mu^2$ in (\ref{fB}) by $\mu_R^2-\mu_L^2=4\mu_A\mu_V$.
Note also that we have assumed that the magnetic field is a perturbation; it 
could be interesting to perform a similar calculation in the case of a strong magnetic field. 

We defer speculating about possible phenomenological consequences of the
vortical and magnetic contributions to the drag force that we have found in (\ref{fO}) and (\ref{fB})
to Section~\ref{sec:Phenomenology}.
We note, however, that both effects are small both because they are 
proportional to $\mu_A$, which arises only due
to topological fluctuations in the plasma, and because they are proportional to $\mu_V/T$ (chiral
magnetic drag force) or  $\mu_V^2/T^2$ (chiral vortical drag force).
This suppression is less severe in lower energy heavy ion collisions in which the
quark number chemical potential is higher, but of course the lower the collision energy
the rarer it is to produce heavy quarks.

\subsection{Contributions to the Drag Force arising from the Gravitational Anomaly}

Before we continue, we introduce a further generalization.
The $T^2$ term in the anomalous axial current in (\ref{ce})  raises the 
question of whether there is an analogue of this term in the drag force. 
Such a term could be substantially greater in magnitude in
any phenomenological consideration, since $T^2$ is
greater than $\mu_V^2$ in heavy ion collisions, and
is much greater than $\mu_A^2$.
The $T^2$ contribution to $J_A$ in (\ref{ce}) is connected to the presence
of a gravitational Chern-Simons term in the bulk metric (see e.g.~Ref.~\cite{Landsteiner:2011cp}),
namely a contribution to the bulk action that we have neglected up to this point
that takes the form
\begin{eqnarray}
\delta S=-\frac{\kappa_g}{16\pi G_5}\int~d^5x\sqrt{-G}\,\epsilon^{MNPQR}A_M R^A_{BNP}R^B_{AQR}\ .
\end{eqnarray}
Here, $\kappa_g$ fixes the coefficient in front of the gravitational axial anomaly in the boundary 
theory. If, following Ref.~\cite{Landsteiner:2011iq}, we add a single chiral fermion transforming under $U(1)_{L}$ to the holographic theory, we then have
\begin{equation}
\kappa_g=\frac{\kappa}{24}\,.
\label{eq:kappacomparison}
\end{equation}
We will use this expression when we make estimates. The value of the ratio (\ref{eq:kappacomparison}) for the axial quark number current in two-flavor QCD is $\kappa_g/\kappa=3/20$ while in the three-flavor case it is $\kappa_g/\kappa=3/16$. 
Introducing $\kappa_g\neq 0$ adds two new terms to the expression for $j_\sigma$ in (\ref{eq:jsigma}) that appears in the bulk metric (\ref{metric}), one in the direction of the fluid angular momentum and one in the direction of the external magnetic field.
Working to zeroth order in gradients and in each case keeping the lowest two nonzero
terms in the expansion in $\mu/T$ we find 
\begin{eqnarray}
j^B_{g,\sigma}&=&\kappa_g B_\sigma \Biggl(\frac{8 \pi ^4 T^4 }{r^6}\notag\\
&~&-\frac{4 \mu ^2 \left(r^2-\pi^2 T^2\right)^2}{\pi^2T^2r^8} \left(\left(r^2+\frac{3\pi^2 T^2}{2}\right)+\frac{r^4\left(r^2+\pi^2T^2\right)}{\pi^2 T^2\left(r^2-\pi^2T^2\right)}\log\left(\frac{r^2}{r^2+\pi ^2 T^2}\right)\right)+{\cal{O}}(\mu^4)\Biggr)\notag\\
j^\ell_{g,\sigma}&=&\kappa_g\ell_\sigma \frac{4 \mu  \left(2 \pi ^2 r^6 T^2-\pi ^4 r^4 T^4+2 r^4 \left(r^4-\pi ^4 T^4\right)\log\left(\frac{r^2}{r^2+\pi ^2 T^2}\right)+3 \pi ^8 T^8\right)}{\pi ^2 r^8 T^2} \notag\\
&~&\qquad-\kappa_g\ell_\sigma\frac{32\sqrt{2}\mu ^3\pi^2T^2}{27 r^{6}} \Bigg(\frac{3\left(6 r^6-3 \pi ^4 r^2 T^4-\pi ^6 T^6\right)}{\pi^6T^6}\log\left(\frac{r^2}{r^2+\pi^2 T^2}\right)\notag\\
&~&\qquad\qquad+\frac{1}{ r^4 \pi^4T^4} \left(18 r^8-9 \pi ^2 r^6
   T^2+\pi ^4 r^4 T^4-18 \pi ^6 r^2 T^6+8 \pi ^8 T^8\right)\Bigg)+{\cal O}(\mu^3)\ .
\end{eqnarray}
Note that $\kappa_g$ flips sign for left-handed quarks, which means that in a theory
in which there are both left- and right-handed quarks the term in $j^B_{g,\sigma}$
that is of order $\kappa_g T^4$ will cancel, and we drop it henceforth.
This means that the $j^B_{g,\sigma}$ is only a quantitative correction to
the $j^B_\sigma$ in (\ref{eq:jBsigma}) that, given (\ref{eq:kappacomparison}), is  small in magnitude.
In contrast, because $j^\ell_{g,\sigma}$ includes a term that is linear in $\mu$ it can
be substantially greater than the $j^\ell_\sigma$ term in (\ref{eq:jsigma}), as anticipated.

Completing the calculation, for the gravitational axial anomaly contribution to the drag force we find 
two terms, one in the $\vec B$ direction and one in the $\vec \ell$ direction, given by
\begin{eqnarray}
\label{fg}
(f^B_g)^\mu_{1,2}&=&-\frac{2\kappa_g}{\kappa}(f^B)^\mu_{1,2}+\frac{\sqrt{\lambda}\kappa_g\mu^2}{3\pi^3s^3T^2\gamma}\Bigl(5s B^\mu+(B\cdot w)(-2u
^\mu+3s w^\mu)\Bigr)\notag\\
(f^\ell_g)^\mu_{1,1}&=&-\frac{\sqrt{\lambda}}{2\pi}\frac{4\kappa_g\mu s^2}{\gamma}\left(C_B\left(\pi T\sqrt{-s}\right)+\frac{3}{s^4}\right)\Bigl(\ell^\mu+(\ell\cdot w)w^\mu\Bigr)\, , 
\end{eqnarray}
where we have only kept the lowest order terms in $\mu$ that are nonzero in each case, and in particular
have dropped the ${\cal O}(\mu^3)$ term in $f^\ell_g$ that could be obtained from
the ${\cal O}(\mu^3)$ term in $j^\ell_{g,\sigma}$.

The new contribution to the chiral
magnetic drag force does not introduce any qualitative change.  The new
contribution to the chiral vortical drag force, however, is linear in $\mu/T$ 
whereas what we had found previously in (\ref{fO}) was of order $(\mu/T)^3$,
meaning that the new contribution (\ref{fg}) is the leading one.
At present no observational evidence of the $T^2$ term in (\ref{ce}) has been
reported (except on the lattice, see Refs.~\cite{Braguta:2013loa,Braguta:2014gea,Buividovich:2013hza})). So, if it is ever possible
to see the chiral vortical drag force, in which (\ref{fg}) is the leading contribution,
this could be a way to probe the physics of the gravitational axial anomaly
in the gauge theory.


\section{The Dissipationless Character of the Chiral Magnetic and Vortical Effects}
\label{sec:Dissipation}

It was pointed out some time ago~\cite{Kharzeev:2011ds} using a symmetry
argument that the charge currents induced by the chiral magnetic and chiral vortical effects
must be dissipationless.  The argument 
in its essence is that these chiral effects are time-reversal invariant,
and so cannot result in the production of entropy.
In this Section we show that our results for the chiral drag
force on a heavy quark provide an example of an explicit
calculation that confirms the dissipationless character of
the CME and CVE.  We do
so by finding a particular setting in which we can place
a heavy quark at rest within momentum and charge currents induced
by either the CME or the CVE, much like placing a defect in a current-carrying
wire or placing a rock in a flowing stream, and see that
in the setting that we construct
the CME- or CVE-induced current flows past the 
heavy quark defect without exerting any drag force upon it.

Throughout this Section we shall largely neglect effects arising from the
gravitational anomaly, in effect setting $\kappa_g=0$.
In order to
achieve our goals in this Section, we need not include
spatial or temporal gradients in the fluid other than vorticity.
Gradients can
also be added, again at the expense of adding complication,
and again without affecting the conclusion.

We shall make our argument in three steps.
First, we begin by noting that a heavy quark that is at rest
in the local fluid rest frame feels a chiral drag force
in a chiral fluid, in the presence of either a magnetic
field $\vec B$ or a fluid vorticity $\vec \Omega$.
The local fluid rest frame is the frame in which
the fluid momentum $T^{0i}$ vanishes at the
location of the heavy quark.  If the heavy quark
is at rest in this frame, then in this frame $u^\alpha=w^\alpha=(1,\vec 0)$,
$\gamma=1$ and $s=-1$ and the leading contributions
to the chiral drag force due to $\vec B$ and $\vec \Omega$
that we have calculated in (\ref{fB}) and (\ref{fO}) simplify to
\begin{equation}
\vec{f}^{B,\ell}_{\rm rest}=-\frac{\kappa\sqrt{\lambda} }{2\pi^3}\frac{\mu^2}{ T^2}\vec{B}-\frac{2\kappa\sqrt{\lambda}}{3\pi^3} \,\frac{\mu^3}{T^2}\,\vec{\Omega}\ ,
\label{ForceAtRest}
\end{equation}
where we have used the fact that $C_B(\pi T)=1$.
%
%
%
Recall that $\mu$ here represents $\mu_R$.  When $\mu_L$ and $\mu_R$ are both nonzero, 
$\mu^2$ is replaced by $4\mu_A\mu_V$ and $\mu^3$ is replaced by $6\mu_A \mu_V^2 + 2 \mu_A^3$. 
It is striking that a heavy quark at rest in a stationary chiral fluid with $\mu_A>0$  ($\mu_A<0$)
feels a force exerted on it by the chiral fluid in a direction antiparallel to (parallel to) the direction of a magnetic field
or fluid vorticity.  (Recall that $\kappa=-\frac{1}{4\sqrt{3}}$ is negative for $\mu=\mu_R$, and recall that $\vec f$ is the 
force that some external agent must exert to keep the heavy quark at rest while $-\vec f$ is the force
that the heavy quark feels from the fluid.)

Second, we ask what we expect will happen if we release the heavy quark at rest
and allow it to move under the action of the force $-\vec f$, with $\vec f$ as in (\ref{ForceAtRest}).  The heavy quark
will start to move through the fluid with some initially increasing velocity $\vec{v}$.
As long as $\vec{v}$ is small in magnitude we can neglect the resulting
change in the chiral drag force (\ref{ForceAtRest}), but because the
heavy quark is now moving it will feel an ordinary drag force (\ref{OrdinaryDrag})
(as calculated in Refs.~\cite{Herzog:2006gh,Gubser:2006bz}) in addition, which for small $v$ is given by
\begin{equation}
\vec{f}_{\rm drag}=\frac{\sqrt{\lambda}}{2\pi} \pi^2 T^2 \vec{v}\ .
\label{ConventionalDrag}
\end{equation}
This means that the heavy quark will accelerate until it reaches
a terminal velocity 
\begin{equation}
\vec{v}_{\rm terminal} =  \kappa \frac{\mu^2\vec B}{(\pi T)^4}  + \frac{4\kappa}{3} \frac{\mu^3 \vec\Omega}{(\pi T)^4} 
\label{TerminalVelocity}
\end{equation}
at which it experiences no further net force.
%
%
We therefore reach the remarkable conclusion that in the local fluid rest frame
a heavy quark at rest feels a force (\ref{ForceAtRest}) while
a heavy quark moving through the fluid with velocity $\vec{v}_{\rm terminal}$ feels no 
force.\footnote{Note that if there were a chemical potential for heavy quarks, meaning
an excess density of heavy quarks compared to the density of heavy antiquarks, then once
all the heavy quarks and antiquarks are moving at $v_{\rm  terminal}$ there would be a resulting
heavy quark current in the local fluid rest frame.}

Third, we boost by a velocity $\vec{v}_{\rm terminal}$
to a new frame in which the heavy quark 
that was moving with velocity $\vec{v}_{\rm terminal}$ in the original frame is now
at rest.  In this new frame, the fluid
at the location of the heavy quark is
flowing with velocity $-\vec{v}_{\rm terminal}$, the heavy quark is at rest, and the heavy
quark feels no force.  The fluid momentum current is proportional to $\kappa$ and is entirely due to the CME and CVE, 
and the heavy quark placed in the flowing fluid feels no
force.  This is an explicit demonstration of the dissipationless character
of the chiral magnetic and vortical effects.

In the remainder of this Section, we will generalize this conclusion in two steps.
First, we will show that the frame in which a heavy quark at rest in the
flowing fluid feels no force is, in fact, the frame in which the
local entropy current
vanishes.  We will then use holography to demonstrate that this conclusion
generalizes.

The full expressions for the charge current $J^\mu$ and the entropy current including
the chiral magnetic and chiral vortical effects are~\cite{Vilenkin:1980fu,Son:2009tf,Sadofyev:2010is,Landsteiner:2011cp}
\begin{eqnarray}
&~&J_\mu=nu_\mu + C\left(\mu-\frac{1}{2}\frac{n\mu^2}{\varepsilon+P}\right)B_\mu+\frac{1}{2}C\left(\mu^2-\frac{2}{3}\frac{n\mu^3}{\varepsilon+P}\right)\ell_\mu\,,
\label{FullChargeCurrent}\\
&~&s_\mu=su_\mu-\frac{1}{2}\frac{C\mu^2s}{\varepsilon+P}B_\mu-\frac{1}{3}\frac{C\mu^3s}{\varepsilon +P}\ell_\mu\,,\label{FullEntropyCurrent}
\end{eqnarray}
where $n$ is the number density of right- or left-handed quarks and
$s$ is the ordinary entropy density. In these expressions, $C$ is the coefficient of the chiral anomaly,
$\partial^\mu J_\mu = C  \vec{E}\cdot \vec{B}$, and is given by 
$C=1/(4\pi^2)$ in a $U(1)$ gauge theory as in (\ref{ce}) while here, in ${\cal N}=4$ SYM theory, we have 
\begin{equation}
C \equiv - \frac{N_c^2 \kappa}{\pi^2} = \frac{N_c^2}{4 \pi^2 \sqrt{3}}\,.
\label{CDefinition}
\end{equation}
%
%
%
We see that when the fluid is at rest there are nonvanishing charge and entropy currents.
However, if we boost to a new frame in which the new (primed) fluid velocity is related to
the original $u^\mu$ by
\begin{equation}
u'_\mu=u_\mu-\frac{1}{2}\frac{C\mu^2}{\varepsilon+P}B_\mu-\frac{1}{3}\frac{C\mu^3}{\varepsilon+P}\ell_\mu
\label{BoostU}
\end{equation}
then in this frame the charge and entropy currents are given by
\begin{eqnarray}
&~&J'_\mu=n u'_\mu + C\mu B_\mu+\frac{1}{2}C \mu^2 \ell_\mu\ ,
\\
&~&s'_\mu=s u'_\mu \ ,
\end{eqnarray}
meaning that we have boosted to a frame in which $u'_\mu=(1,\vec{0})$ means
no entropy current.\footnote{In the presence of the gravitational anomaly, {\it i.e.}~if $\kappa_g\neq 0$, the entropy current 
(\ref{FullEntropyCurrent}) includes additional terms and, consequently, 
does not vanish in the ``no drag frame''~\cite{misha_yee}.}
Note, of course, that when $u'_\mu=(1,\vec{0})$ there is
a nonzero momentum current, $T^{0i}\neq 0$;  the fluid momentum vanishes when $u_\mu=(1,\vec{0})$.
In the nonrelativistic limit, the boost velocity needed to accomplish (\ref{BoostU}), namely
to boost from the local fluid rest frame to the local entropy rest frame, is
\begin{equation}
\label{term_vel_hydro}
\vec v_{\rm boost}=-\frac{1}{2}\frac{C}{\varepsilon+P}\left(\mu^2\vec B+\frac{4}{3}\mu^3\vec \Omega\right)
\end{equation}
where we have used the nonrelativistic approximation $\vec l\simeq2\vec\Omega$.  Upon using (\ref{CDefinition}), noting
that $\varepsilon+P=N_c^2 \pi^2 T^4/2$, and comparing to (\ref{TerminalVelocity})  we see that
\begin{equation}
v_{\rm boost}=v_{\rm terminal} \,.
\end{equation}
We conclude that the frame in which a heavy quark at rest in the flowing
fluid feels no force is in fact the local entropy rest frame.

The generality of this conclusion will become apparent after it is recast
holographically, in terms of the dual gravitational description of the fluid.
In particular, upon so doing we will see that 
although in the derivation above
we only used the contributions to the chiral magnetic and chiral
vortical drag force and to the ordinary drag force
that arise at the lowest nontrivial order in $\mu/T$ in each case, the conclusion
that we have reached in fact holds to any order in $\mu/T$.

The authors of Refs.~\cite{Loganayagam:2011mu,Chapman:2012my,Eling:2012xa,Megias:2013xla} have shown
that 
in the holographic description of a flowing strongly coupled plasma with $\kappa_g=0$ with
the 4+1-dimensional metric (\ref{metric}) the local entropy rest frame is
the frame in which the metric function $j_\sigma$ in (\ref{metric}) vanishes
at $r=r_h$, where $r_h$ is the location of the outer horizon, namely
the largest solution of $f(r)=0$, with $f(r)$ also being a metric function
in (\ref{metric}).  For the case of a static fluid with $\mu=0$, $r_h$ is given
simply by $r_h=\pi T$.
In a static fluid with  $\mu\neq 0$, we have instead $r_h=r_+$
where $r_+$ is given in terms of $\mu$ and $T$ in (\ref{rplus}).
%
%
%
%
%
%
We must therefore compute the force on a heavy quark {\it at rest} in the frame in which
$j(r_h)=0$.  This computation turns out to be sufficiently tractable that
we can push it far enough without expanding in $\mu/T$.  We shall include
a nonzero magnetic field and vorticity.  As throughout this Section, we neglect
the gravitational anomaly and fluid gradients.
In this case, $j_\sigma$ takes the form
\begin{eqnarray}
j_\sigma= 
W^B(r)B_\sigma + W^\ell(r)\ell_\sigma 
\label{jsigmaExpression}
\end{eqnarray}
where the effects of the chiral anomaly enter through the
functions $W^B(r)$ and $W^\ell(r)$ whose form we obtained
explicitly to leading nontrivial order in $\mu/T$ in (\ref{eq:jBsigma}) and (\ref{eq:jsigma}), respectively.
%
%
%
The force on the heavy quark at rest can be calculated following the procedures
developed in previous Sections, and one finds
\begin{eqnarray}
\label{frame_drag}
\vec{f}=-\frac{\sqrt{\lambda}}{2\pi}\left( 
r_h^2W^B(r_h)\vec{B}+r_h^2W^\ell(r_h)\vec{\ell}~\right).
\end{eqnarray}
It can be also verified that if one expands this answer for the anomalous drag force 
in powers of $\mu/T$, it does indeed coincide
with (\ref{fB}) and (\ref{fO}).
If we balance this force against the drag force on a slowly moving quark,
which in this general context is given by
\begin{eqnarray}
\vec{f}=\frac{\sqrt{\lambda}}{2\pi}r_h^2\vec{v}\ ,
\label{OrdinaryDragGeneral}
\end{eqnarray}
we see that the terminal velocity of the heavy quark is given by
\begin{eqnarray}
\label{term_vel_hol}
\vec{v}_{\rm terminal}=\left( 
W^B(r_h)\vec{B}+W^\ell(r_h)\vec{\ell}~\right).
\end{eqnarray}
Comparing this result with (\ref{jsigmaExpression}), we see that in the local entropy rest frame 
in which $j_\sigma(r_h)=0$, the terminal velocity vanishes: $v_{\rm terminal}=0$.  This
means that in the local entropy rest frame, a heavy quark at rest feels no force. 
Note that in obtaining this result via this holographic
calculation we did not need to expand $W^B(r)$ or $W^\ell(r)$ in powers of
$\mu/T$.  This means that the result (\ref{term_vel_hol}) is valid to all orders in $\mu/T$.
Of course, in (\ref{eq:jBsigma}) and (\ref{eq:jsigma}) we have explicit expressions
for $W^B(r_h)$ and $W^\ell(r_H)$ only to lowest nontrivial order in $\mu/T$; at
higher orders, $W^B(r_h)$ and $W^\ell(r_H)$ receive corrections but the form
of the expression (\ref{term_vel_hol}) for the terminal velocity of the heavy quark
remains unchanged.

Note also that although we have not given the resulting more complicated
expressions here, if one adds
the effects of fluid gradients into the expressions
for $j_\sigma$, the entropy current, and the drag force, while keeping $\kappa_g$ set to zero, the conclusion that
a heavy quark at rest in the local entropy rest frame feels no force remains
unchanged.

Let us close this section by restating this general result
in its simplest form.  In a strongly coupled fluid in which $\mu_R > \mu_L$, in the presence
of a magnetic field $\vec B$ and for simplicity in the absence
of any fluid gradients including vorticity, we 
have analyzed the chiral drag force on heavy
quarks with two different velocities:
\begin{itemize}
\item
If the first heavy quark is at rest in the local fluid rest frame,
meaning that it sees around it a fluid with no momentum flow,
this heavy quark feels the fluid around
it exerting a chiral drag force on it pushing it toward the direction of $-\vec B$.
It also sees around it a charge current in the direction of $\vec B$
and an entropy current in the opposite, $-\vec{B}$, direction.
(All signs are reversed if instead the fluid has $\mu_L>\mu_R$.)
\item
If a second heavy quark is released and allowed to accelerate
under the influence of the chiral drag force, it does so until
it reaches a terminal velocity in the $-\vec B$ direction.  
If we now boost to the rest frame of {\it this} heavy quark,
we find a heavy quark that feels no force.  This heavy quark sees a charge
current in the $\vec B$ direction that is larger in magnitude than that
seen by the first heavy quark.  This heavy quark also
sees the fluid around it flowing, with fluid momentum 
in the $\vec B$ direction.
(Again, all signs are reversed if instead the fluid has $\mu_L>\mu_R$.)
\end{itemize}
The second heavy quark provides an example of a 
``defect'' at rest in a flowing chiral fluid, with both
fluid momentum and charge flowing past it, and yet
feels no force.  Thus, it provides an explicit example
of the dissipationless character of the chiral magnetic
and chiral vortical effects.  

This becomes particularly important now that
the chiral magnetic effect has been seen~\cite{Li:2014bha} in a 
condensed matter system,
namely the Dirac semi-metal ZrTe$_5$, 
since we can assume that
the zirconium pentatelluride crystals used in the
experiment contain some defects.

\section{Outlook and Potential Phenomenological Consequences}
\label{sec:Phenomenology}

In Section~\ref{sec:Gradients}, we obtained the contributions to the drag force on a heavy
quark that are second order in $\mu/T$ and first order in fluid gradients in
a fluid in which $\mu_L=\mu_R$, meaning that there are no contributions
to the drag force coming from the chiral anomaly.  These results generalize
the $\mu=0$ results of Ref.~\cite{Lekaveckas:2013lha} to nonzero $\mu$.
In Ref.~\cite{Lekaveckas:2013lha}, the analytic results for the
contributions to the drag force that are first order in gradients
were used to obtain an understanding of a variety of curious features
of the drag force on a heavy quark caught between two colliding sheets
of energy density, features that were discovered numerically
in the holographic calculation of Ref.~\cite{Chesler:2013cqa}. 
We look forward to seeing our results from Section~\ref{sec:Gradients}
used similarly, in conjunction with some future calculation of the collision
of two sheets of energy and charge density. 

We have calculated the anomalous contributions to the drag force on a heavy
quark in a chiral plasma to lowest nontrivial
order in $\mu/T$ in Section 4.  
These contributions to the drag force change sign
in a chiral plasma with $\mu_R>\mu_L$ relative to that with $\mu_L > \mu_R$.
This parity-odd symmetry makes the chiral drag force novel, qualitatively
different from all contributions to the drag force on heavy
quarks that have been calculated previously.  Its unique
features --- namely that it is a force that pushes all heavy quarks and
antiquarks in the same direction, either parallel to or antiparallel to
the magnetic field vector $\vec{B}$ or the fluid vorticity vector $\vec{\Omega}$, with the direction
of the force being opposite in 
a chiral plasma with $\mu_R>\mu_L$ relative to that with $\mu_L > \mu_R$ --- make
it detectable at least in principle via suitable  observables defined for this purpose.  In practice
the effect is likely to be small in quantitative terms.
We shall provide a rough estimate of the magnitude of the 
chiral drag force, before describing a possible observable.

The full leading order expressions for the chiral drag force are given in
Section 4 for a quark moving through a chiral plasma with any velocity.  
Here we recapitulate our result for the anomalous force on a heavy
quark at rest  in the local fluid rest frame 
\begin{equation}
\frac{2\pi}{\sqrt{\lambda}}\vec{f}=-\frac{(3\kappa-16\kappa_g)\mu^2}{3\pi^2 T^2}\vec{B}-32\kappa_g\mu\vec{\Omega}-\frac{4\mu^3(\kappa-16\kappa_g)}{3\pi^2 T^2}\vec{\Omega}\ ,
\label{ForceAtRest2}
\end{equation}
where we have restored the effects due to the gravitational anomaly, proportional
to $\kappa_g$, that we dropped in (\ref{ForceAtRest}).
We can compare this result for the chiral drag force on a heavy quark
with $v=0$ to the result for a heavy quark in the $v\rightarrow 1$ limit, where we 
find
\begin{equation}
\frac{2\pi}{\sqrt{\lambda}}\vec{f}=-\frac{4(\kappa-2\kappa_g)\mu^2}{3\pi^2 T^2}(\vec{B}\cdot\vec{v})\vec{v}-\frac{32\kappa_g\mu}{3}(\vec{\Omega}\cdot \vec{v})\vec{v}-\frac{4\mu^3(3\kappa-40\kappa_g)}{9\pi^2 T^2}(\vec{\Omega}\cdot \vec{v})\vec{v}\ .
\label{ForceUltrarelativistic}
\end{equation}
There are certainly differences between (\ref{ForceUltrarelativistic}) and (\ref{ForceAtRest2}), for example
the fact that in the ultrarelativistic limit the force is in the $\vec v$ direction rather than in the 
$\vec B$ or $\vec \Omega$ direction, although its magnitude is greatest when $\vec v$
is parallel or antiparallel to $\vec B$ or $\vec \Omega$.
The principal message coming from (\ref{ForceUltrarelativistic}), however, is that the anomalous
drag force on a heavy quark with $v\rightarrow 1$ is comparable in magnitude to that on
a heavy quark at rest.  All heavy quarks and antiquarks feel a chiral drag force whose
magnitude is only weakly dependent on their velocity.
Given that, it seems clear that the fractional effects of
the chiral drag force are largest when $v\sim 0$ rather than
when $v\rightarrow 1$.  Although in principle there are anomalous forces on
acting on the ultrarelativistic b-quarks that become b-jets, it seems a better
bet to focus on b-quarks that are almost at rest relative to the fluid. We shall
see that even in this case the effects are small.

Let us consider a heavy quark with mass $m$ that is initially at rest in the local fluid
rest frame in a chiral plasma in the presence of a magnetic field $\vec B$.
This quark feels a chiral drag force
\begin{equation}
\vec{f}^B = - \kappa\sqrt{\lambda}\left(1-\frac{16\kappa_g}{3\kappa}\right) \frac{\mu^2}{2\pi^3 T^2}\vec{B}\ .
\end{equation}
As we described in Section 5, this heavy quark starts
to move due to this force and as it moves  it begins to feel
the standard drag force (\ref{ConventionalDrag}) also, meaning that the heavy quark accelerates 
until
it is moving with a terminal velocity 
\begin{equation}
\vec{v}_{\rm terminal}=\kappa\left(1-\frac{16\kappa_g}{3\kappa}\right) \frac{\mu^2} {(\pi T)^4} \vec{B}\,.
\end{equation}
How
long does it take for the heavy quark to accelerate from rest to a velocity that is close to $v_{\rm terminal}$?
This timescale is parametrically of order $m |\vec{v}_{\rm terminal}|/|\vec{f}^B| \sim 2\pi m /(\sqrt{\lambda}(\pi T)^2)$,
meaning that for charm (bottom) quarks it is of order 1 fm/$c$ (a few fm/$c$).  For our rough
purposes we can therefore estimate that the chiral drag force gives all heavy quarks 
a momentum that is around $m\, \vec{v}_{\rm terminal}$.  Bottom quarks produced at
rest in the fluid therefore pick up a contribution to their momenta arising from the
chiral anomaly that is of order
\begin{eqnarray}
\label{FinalNumericalEstimate}
\vec{p}_{\rm terminal}&\equiv& m_b\, \vec{v}_{\rm terminal}\nonumber\\
& = & m_b \,\kappa\left(1-\frac{16\kappa_g}{3\kappa}\right) \frac{\mu^2 \vec{B} } {(\pi T)^4}\nonumber\\
& = & - 0.449\, m_b \frac{\mu_V \mu_A \vec{B} } {(\pi T)^4}\nonumber\\
& \simeq & - 3~{\rm MeV}\, \frac{m_b}{4.2~{\rm GeV}}\,\frac{\mu_V}{0.1~{\rm GeV}} \,\frac{\mu_A}{0.1~{\rm GeV}} \,\frac{\vec{B} }{(0.1~{\rm GeV})^2}\left(\frac{0.5~{\rm GeV}} {\pi T} \right)^4\,,
\end{eqnarray}
where we have used (\ref{KappaValue}) and (\ref{eq:kappacomparison}), and replaced $\mu^2$ by $4\mu_A\mu_V$.
For concreteness, we have used the ${\cal N}=4$ SYM values in evaluating the purely numerical factor $\kappa(1-\frac{16\kappa_g}{3\kappa})$;
in QCD all components of this coefficient --- namely $\kappa$, $\kappa_g/\kappa$ and the 16/3 --- will take on
different values. 
There will certainly also be other places where our ${\cal N}=4$ SYM calculation 
differs from QCD by factors of order unity.  Note, finally, that we have throughout used
units in which a factor of $e$ has been absorbed into $\vec{B}$ and $\vec{E}$. Restoring
this means replacing $\vec{B}$ by $e\vec{B}$ in (\ref{FinalNumericalEstimate}).

Reading the final expression (\ref{FinalNumericalEstimate}) from left to right, we see that:
\begin{itemize}
\item
The effect is small.
\item
The effect can be enhanced by lowering the collision energy, as doing so increases $\mu_V$, but one should
not lower the collision energy too far since doing so makes heavy quark production rarer and also
reduces the initial temperatures reached in the collision.  Generating the chiral effects we are discussing
requires the presence of quark-gluon plasma.  Recent measurements~\cite{Adamczyk:2014mzf}
indicate that the observables that are sensitive to the chiral magnetic
effect that were seen previously in high energy heavy ion 
collisions~\cite{Abelev:2009ac,Abelev:2009ad,Abelev:2012pa,Adamczyk:2013hsi} persist robustly down 
to collision energies corresponding to baryon chemical potentials $\mu_B>0.3$~GeV,
meaning $\mu_V>0.1$~GeV.
\item
It is difficult to estimate the magnitude of the fluctuations in $\mu_A$ in heavy ion collisions.
The authors of Ref.~\cite{Hirono:2014oda} have recently estimated that $\mu_A$ can locally be as large as 0.1 GeV
in heavy ion collisions at top RHIC energies.  Estimating this more reliably will require
the development of a relativistic viscous chiral magnetohydrodynamic code for heavy
ion collisions. 
\item
In heavy ion collisions with a nonzero impact parameter, a magnetic field is created initially
by the charged spectators.
The conductivity of the quark-gluon plasma slows the decay of the magnetic field, delaying
the decay of the magnetic field in the plasma to long after the spectators are gone. Reliable
estimates here also await a future magnetohydrodynamic analysis of heavy ion collisions,
but perhaps $|\vec {B}|$ can remain as large as $(0.1~{\rm GeV})^2$ for a few fm/$c$~\cite{Tuchin:2013ie,McLerran:2013hla,Tuchin:2013apa,Gursoy:2014aka},
although this is more likely to be an overestimate than an underestimate.
\item
Taking the average temperature seen by the heavy quark 
as $\pi T=0.5$~GeV is on the low side for a 
benchmark value with which to make an estimate, although maybe not unreasonable
for heavy ion collisions with $\mu_V>0.1$~GeV.
\end{itemize}
Our estimates suggest that the chiral magnetic drag force on bottom (charm) quarks and antiquarks pushes
them all to the extent that it gives them all a common momentum that might be 
as large as  $\sim 3$~MeV ($\sim 1$~MeV), although it should be clear that
this estimate is first of all at present very crude and second of all is more likely to be
an overestimate than an underestimate.\footnote{From (\ref{ForceAtRest2}) we see that there is a chance
that the chiral vortical drag force could be larger than the chiral
magnetic drag force since although $\kappa_g\ll \kappa$ 
the chiral vortical drag force is larger by some purely numerical factors and because it is not suppressed by $\mu_V/(\pi T)$.
However, at present the magnitude of $\vec\Omega$ in the droplets of fluid produced
in heavy ion collisions is poorly constrained and we have also
only just seen the first measurements of observables that can receive a contribution
from the chiral vortical effect~\cite{Zhao:2014aja}.  We will therefore not attempt an estimate
of the magnitude of the chiral vortical drag force.} Furthermore, in events in which
there are some regions of the plasma with $\mu_R>\mu_L$ and other regions with $\mu_R<\mu_L$,
some heavy quarks will be pushed in the direction antiparallel to $\vec B$ while others
will be pushed parallel to $\vec B$, reducing the net effect in the event as a whole.

Although the effects of the chiral drag force are small,
we do not wish to underestimate the ingenuity of our
experimentalist colleagues.  Perhaps they can devise sufficiently sensitive correlation
observables to see the small effects of the chiral drag force on heavy quarks. For example,
it is worth constructing event-by-event observables that can see whether the $B$ and $D$ mesons in
an event have picked up a net momentum perpendicular to the
reaction plane in the  ``downward'' (``upward'') direction
in those events in which the
chiral magnetic effect has resulted in an electric current ``upwards'' (``downwards'')
with positive light hadrons pushed ``upwards'' (``downwards'') and
negative light hadrons pushed ``downwards'' (``upwards''). 
A correlation observable like this uses the CME current
to define the direction in which the heavy quarks should be pushed by the chiral drag force,
and then checks whether a net push on all the heavy quarks and antiquarks in the
event in this direction is seen.  A nice feature of such a correlation observable is that
the observable effects of both the CME current and the chiral drag force should
each be equally suppressed by the partial cancellation between regions of the plasma 
with $\mu_R>\mu_L$
and regions with $\mu_R < \mu_L$.

Given the estimates that we have made, it is certainly our impression that the chiral
drag force is of interest principally from a theoretical perspective, rather than as a phenomenon
that experimentalists should expect to observe in heavy ion collisions.  
Even if they
are only observable in principle, though, it is remarkable to see effects due to the chiral anomaly
giving all the heavy quarks and antiquarks in a chiral plasma a kick.  
In Section~\ref{sec:Dissipation} we have also used our calculation of the chiral drag force to provide
explicit evidence for the dissipationless character of the chiral magnetic and chiral
vortical effects, showing that the currents that they describe can flow around defects without
any dissipation.

\begin{acknowledgments}

We would like to thank M.~Lekaveckas for his contributions
during the early stages of this work, as we were setting up the
calculation of the drag force in a plasma with both gradients and chemical potential.
We would like to thank M.~Stephanov and H.-U.~Yee for 
comments that have helped us to improve our paper and we would also like
to thank 
D.~Kharzeev, H.~Liu, A.~M.~Polyakov, D.~T.~Son, Y.~Yin and V.~I.~Zakharov
for helpful discussions. 
This work was supported by the U.S. Department of Energy
under Contract Number DE-SC0011090.  
AS is grateful
for travel support from RFBR grant 14-02-01185A during
his visits to Russia.

\end{acknowledgments}

\end{document}